\documentstyle[preprint,aps]{revtex}
\newcommand{\bce}{\begin{center}}
\newcommand{\ece}{\end{center}}
\newcommand{\be}{\begin{equation}}
\newcommand{\ee}{\end{equation}}
\newcommand{\bea}{\vspace{0.25cm}\begin{eqnarray}}
\newcommand{\eea}{\end{eqnarray}}

\def\PLA{{Phys. Lett.}  A }
\def\PRL{{Phys. Rev. Lett.} }
\def\PRA{{Phys. Rev.} A }

\begin{document} \draft

\vskip 3cm

\begin{center}
{\bf {\LARGE Double entanglement and quantum cryptography}}
\end{center}

\vspace{ .25cm}
\begin{center}
{M.Genovese \footnote{ \small  genovese@ien.it} and C. Novero} 
\\[0pt]

Istituto Elettrotecnico Nazionale Galileo Ferraris \\[0pt]
Str. delle Cacce 91 \\[0pt]
I-10135 Torino, Italy
\end{center}

\vspace{ 3.5 cm} {\large  Abstract }
\vskip 0.5 cm 
We propose a quantum transmission based on bi-photons which are doubly-entangled both in polarisation and phase.  This scheme finds a natural application in quantum cryptography, where we show that an eventual eavesdropper is bound to introduce a larger error on the quantum communication than for a single entangled bi-photon communication, when he steels the same information.  
\vskip 0.5 cm
\vskip 1.7cm

The recent fast development of quantum states manipulation techniques has led to new technological applications  of quantum mechanics.
Among different applications of quantum mechanics to technology the possibility of transmitting absolutely confidential messages is of the greatest interest. This is due to the possibility of creating a key for encoding and decoding  secret messages by transmitting single quanta between two parties (usually dubbed Alice and Bob).
The underlying principle of quantum key distribution (QKD) is that nature prohibits gaining information on the state of a quantum system without disturbing it (in particular no-cloning theorem guarantees that one cannot generates copies of an arbitrary unknown state). Thus possible eavesdropping by a third party (usually dubbed Eve) can be identified. This is at variance with current methods of public key cryptography, which are based on the supposed, but unproven, classical computational difficulty in solving certain problems,  e.g. factoring  large numbers in prime factors. Furthermore, a quantum computer would efficiently solve these problems \cite{shor} breaking these kind of classical cryptographic protocols. From this it arises the great interest in understanding and developing secure quantum cryptographic schemes. 

Since the original proposal of quantum cryptography \cite{Wiesner}, many different protocols for this kind of transmission have been suggested \cite{cryp,BB84,Ekert,B92,oth}.

For example, in Ekert's  protocol \cite{Ekert} entangled pairs are used. Both Alice and Bob receive one particle of the entangled pair. Then they perform a measurement choosing  among at least three different  selections.    Alice and Bob communicate on a classical channel the bases they have used: if measurements were performed in the same basis, they are perfectly correlated and can be used for generating the secret key. The other measurements can be used for a test of Bell inequalities. If a third party, Eve, tries  to eavesdrop, she inevitably affects the entanglement between the two particles leading to a reduction of the violation of the Bell inequalities, which allows  Alice and Bob to recognise the presence of the spy.

In the BB84 scheme \cite{BB84} single states are transmitted from Alice to Bob, preparing them at random in four partly orthogonal states (for example, using photons, in polarisation states at $0^o$ and $90^o$, $45^o$  and $135^o$). Bob selects the bases for the measurement  at random  too. Then Alice and Bob communicate on a classical channel the bases they have used (but not the results of course): when they have used the same basis Bob knows Alice's result and vice versa and they can build a key.
If  Eve tries to intercept the message, she inevitably introduces errors, which Alice and Bob can detect by comparing a subsample of the generated key using the classic channel (which in these schemes is supposed to be subject to eavesdropping, but not alterable).

Many different experiments have been realised using the former schemes, demonstrating the feasibility of QKD up to a distance of many kilometers \cite{lontano} both in air and in fibre. 

All of them are based on transmission of single photon states or weak coherent states, where the alphabet is based either on photon polarisation or on photon phase. It must be noticed that in the case of weak coherent states the transmission can, in principle, be unsafe for sometimes the pulses necessarily contain more than one photon leaving the possibility to an eavesdropper of using these events for gaining information about the key without introducing any extra error \cite{durt}. The use of single photon sources closes this potential security loophole.

General theorems have been demonstrated (mainly for BB84 protocol) which guarantees the security of quantum cryptography in an ideal case \cite{NC,Lo}, albeit no complete demonstration for every  conceivable attack exists \footnote{It must be noticed that a completely general security demonstration could even be impossible in presence of trojan horse attacks, i.e. when the eavesdropper can introduce some unwanted material inside Bob lab using unused (for cryptographic transmission) degrees of freedom of the quantum states. The demonstration of security against trojan horse attacks in ref. \cite{Lo} does note really solve the problem because it is based on teleportation, which requires the use of EPR pairs shared with an insecure area where they can be subject to Eve's manipulation.}.    However, real experimental schemes suffer of huge losses and the application of these theorems is limited. Therefore, it is mandatory the search for strategies which restrict the information potentially obtainable by eavesdropping on real channels.

In this paper we propose the realisation of  "double entanglement" on a single photon pair to be used for quantum communication. More in details, the bi-photon pair is entangled both in polarisation and in phase, allowing a larger bit transmission for every pair and making more difficult a successful eavesdropping.

The scheme for producing such an entanglement is relatively simple: for example it can be realised placing on the pump beam  a Mach-Zender interferometer (whose path length difference is large compared to the pump pulse length) before the non-linear system where a polarisation entangled pair is generated. The pump photon can thus follow the short or the long path originating the superposition 
\cite{Gisen}:
\be
\vert \Psi_p \rangle = { 1 \over \sqrt{2}} \left [ \vert s \rangle + e^{i \phi} \vert l \rangle \right ]
\ee
where  $\vert s \rangle $ and $ \vert l \rangle $ denote the photon which has followed the short and the long path respectively and $\phi$ the phase difference between the two paths.

Then the pump photon creates a photon pair entangled in polarisation  by parametric down conversion or in a type II crystal \cite{typeII} either in two sequential type I crystals \cite{nos} (see fig.1). The second solution presents some advantages for it does not have the problem of different propagation of idler and signal inside the crystal due to different polarisation in a birefringent medium \cite{keller}, furthermore every Bell state can be easily obtained. Finally, using two type I crystals and tuning the pump wave length, one can generate an entanglement on two different frequencies: in this case one wave length could, for example, be chosen at the maximum of transmission of an optical fibre or air and the other (the one remaining in Alice's laboratory) at the one maximising detection efficiency. Incidentally, it must also be noticed that by using this scheme one could also easily obtain non-maximally entanglement both in phase (using a not $50 \% -50 \%$ beam splitter) and polarisation (attenuating the pump between the two crystals or/and using crystals of different lengths) \cite{nos}: this possibility has  relevance for experiments on foundations of quantum mechanics.  
 
Using the two type I crystals scheme, the final bi-photon state is:

\be
\vert \Psi \rangle = { 1 \over 2} \left [ \vert s H \rangle \vert s H \rangle + \vert s V \rangle \vert s V \rangle + e^{i \phi} (\vert l V \rangle \vert l V \rangle + \vert l H \rangle \vert l H \rangle ) \right ] 
\label{psi}
\ee 
where $H$ and $V$ denote the horizontal and vertical polarisation respectively, whilst $\vert s \rangle $ and $ \vert l \rangle $ denote a photon created by a pump photon having travelled via the short or the long arm of the interferometer. 

This is the state that will be used for quantum transmission. 

It must be noticed that this state  is invariant in its form changing the polarisation or the phase basis. For example if the basis
\be
\vert \pm \rangle = { 1 \over \sqrt{2}} \left [ \vert H \rangle \pm \vert V \rangle \right ]
\ee
is chosen, the state \ref{psi} can be rewritten as:
\be
\vert \Psi \rangle = { 1 \over 2} \left [ \vert s + \rangle \vert s + \rangle + \vert s - \rangle \vert s - \rangle + e^{i \phi} (\vert l + \rangle \vert l + \rangle + \vert l - \rangle \vert l - \rangle ) \right ] 
\label{psi2}
\ee 

For implementing quantum communication, one photon is sent to Alice, the other to Bob. Both select the photon by its polarisation (for example using a birefringent prism), choosing different bases and then send it to a Mach-Zender interferometer, which introduces exactly the same  difference of travel times, within the coherence time of the down converted photons, through the two arms as the interferometer on the pump (see \cite{Gisen}). Here they can choose different phases for the long arm (see fig.2). The probability for detection in the central time slot \footnote{corresponding to the two indistinguishable situations when the pump photon has followed the short (long) arm of the interferometer and the two down converted photons both the long (short) one (see \cite{Gisen}).}by a given combination of detectors depends on the phases (e.g. $\phi, \tau _A, \psi _B $) of the three interferometers involved in production and detection of the photon pair \cite{Gisen} and on the polarisers settings. Different choices originate different detection bases.   

Therefore, this scheme allows to obtain two bits for each received photon, one related to polarisation, the other to phase.
When Alice and Bob have chosen the same two bases (both for polarisation and phase) they have two correlated outputs, which they can use for generating the key. The other choices can be used for testing the channel (e.g. by means of Bell inequalities in the  Ekert's protocol).
  
In order to identify without error the state, Eve should hit both the bases.
This probability is reduced to $P^2$ respect to $P$ for a single entangled quantum channel.

In order to quantify this statement, let us begin considering  the case where Alice and Bob use the BB84 protocol (Eve produces the pair, keeps a photon which she will measure in one of the two bases and send the other to Bob) and the final key is given by the sum (modulo 2) of the two results obtained by Alice and Bob in the two bases when these have been chosen in the same way. In this case the communication channel is a binary symmetric one and the information on the channel is given by \cite{dover}
\be
I = 1 + p log_2 p + (1-p) log_2 (1-p) 
\label{I}
\ee
where $p$ is the probability for a correct transmission.

Let us begin considering the simplest case where Eve decides to eavesdrop the photons directed to Bob in one of the possible basis used by Alice and Bob, both for the phase and the polarisation ones. She has a probability $ q_1={ 1 \over 4}$ for mistaking the transmitted bit on a single basis, leading to a probability of having a correct interception on the double entangled system of $ p_2=5/8$, which gives, for Eq. \ref{I}, an information on the Alice-Eve channel $I_{AE} = 0.046$. Besides, Eve introduces a fraction of errors $q_{AB}=15/32$ on the Alice - Bob channel, leading to an information on the Alice-Bob channel $I_{AB} =0.0028$.

If this eavesdropping scheme would have been applied to a single entangled channel, Eve would have got a $q_1$ error rate leading to $I_{AE} = 0.189$ and would have produced a $3/8$  error rate on the Alice - Bob channel, corresponding to $I_{AB} = 0.046$.

If Eve intercepts a fraction $\eta $ of the transmitted photons, she  obtains an information $I_{AE} = 0.189 \eta$ for the single entangled channel and $I_{AE} = 0.046 \eta$ for the double entangled one. In order to obtain the same information she will thus produce an error rate on the Alice-Bob channel $7.7$ larger for the double entangled channel, making by far easier her identification in this case.

Let us then consider the case where Eve chooses for eavesdropping an intermediate basis (dubbed the Breidbart basis) for both the phase and the polarisation ones respect to the bases used by Alice and Bob. This choice does not introduce asymmetric errors, making more difficult the identification of the eavesdropper. The probability for Eve to get a wrong result for a single basis is $ q_1={ 2 - \sqrt{2} \over 4}$, leading to a probability, for our scheme, of having a correct interception of $ p_2=q_1^2 + (1-q_1)^2=1/4$, which gives, for Eq. \ref{I}, $I_{AE} = 0.189$. Furthermore, Eve introduces a fraction of errors $q_{AB}=3/8$ on the Alice - Bob channel, leading to $I_{AB} =0.046$.

On the other hand for the single entanglement, Eve has a $q_1$ error rate leading to $I_{AE} = 0.399$ and produces $1/4$ of error rate on the Alice - Bob channel, with $I_{AB} = 0.189$.

Eavesdropping a fraction $\eta$ of the  photons going to Bob, she  obtains an information $I_{AE} = 0.389 \eta $ for the single entangled channel and $I_{AE} = 0.189 \eta $ for the double entangled one. In order to obtain the same information she shall thus produce an error rate on the Alice-Bob channel $19/6$ larger for the double entangled channel, which, as before, results in a much larger  chance of identifying the eavesdropping in the double entangled case.

This result can also be obtained looking to the case where Alice and Bob adopt an error correction procedure. If they eliminate all the errors and Eve has intercepted a fraction $\eta$ of photons, the upper limit on the information she could eavesdrop is \cite{Hut} $I_{AE} = 0.299 \eta \alpha$ for the single entangled channel, whilst this is reduced to $I_{AE} = 0.118 \eta \alpha$ for the double entangled channel, where $\alpha$ is the reduction factor of the key length during the error correction procedure. 
Therefore, this result shows once again that (even if an exhaustive discussion of the value of $\alpha $ is missing \cite{Hut}) 
the use of the double entangled channel allows a large improvement of the transmission security.  

As a further example, let us consider the effect on a simple implementation of Ekert's protocol, like the one realised in ref. \cite{zei}. In this case Alice and Bob measure their photons each on two bases. One of the bases of Bob and Alice coincides and therefore, when both use this basis, they obtain perfectly correlated results, which are used to build the key. The other results are used for measuring the Wigner inequality:
\be
W=p(\chi,\psi)+p(\psi,\omega)-p(\chi,\omega) \ge 0
\label{wig}
\ee
where $ p(\chi,\psi)$ is the coincidence probability function for the measurement settings $\chi $ and $\psi $ of Alice and Bob respectively.  
This inequality is always satisfied for any local realistic theory, but it is violated in quantum mechanics for an appropriate choice of settings. The maximal violation is $W=-1/8$. If Eve intercepts a fraction of photons, she reduces
the violation of Eq. \ref{wig}. In the implementation of ref. \cite{zei} the detection efficiency of each photon path is  $5 \%$ and the inequality \ref{wig} is measured with a $10 \%$ relative uncertainty. If Eve eavesdrops the photons on the commune basis, she obtains a perfect information of the key for the intercepted photons. However, a $10 \%$ relative uncertainty on the measurement of the Wigner function requires that Eve must intercept a fraction of $6.7 \%$ or smaller of the photons addressed to Bob for remaining undetected. 

If a double entangled channel is used, Eve would affect the value of two Wigner inequalities at the same time, this requires that she reduces (for not being discovered) the intercepted fraction to $4.7 \%$, leading to a reduction of a factor $0.7$ for the eavesdropped information in comparison with the single entangled channel.   
 
Finally, as a last example, let us consider the case where Eve decides to eavesdrop on a generic basis given by a superposition of the basis states 
$\vert s H \rangle $, $\vert l H \rangle $, $\vert \,s V \rangle $ and $\vert l V \rangle $. She chooses at random the basis for the measurement, using a generic SO(4) (SO(2) for the single entanglement) transformation of the previous basis: in this way, on average, no asymmetric error is introduced. After having performed the measurement, Eve passes the photon to Bob exactly in the same state she found it in. In order to understand the effect of such a procedure we have performed a Montecarlo simulation of the eavesdropping, evaluating the errors on the Alice-Bob channel. Our numerical results shows that the errors on the Alice-Bob channel are increased of a factor 1.25 about for the double entangled channel respect to the single entangled one, leading to an easier detection of the eavesdropper in the double entangled channel for this example as well.     
 
A general discussion of security in presence of joint or coherent attacks is beyond the purpose of this work, however, it is evident how the presence of a double entanglement makes much more complicate the use of a translucent interception scheme as, for example, the one described in ref. \cite{M}. Thus one can expect that double entanglement should also be efficient in increasing security against these kinds of eavesdropping. 

In summary, we have shown that the use of states entangled on two (or eventually more) quantum degrees of freedom at the same time allows a safer communication for realistic quantum channels. We have also proposed a scheme for obtaining such a double entanglement, which can be easily realised with a simple modification of present experiments. 

\vskip 1cm
{\bf Acknowledgements}

\noindent We would like to acknowledge support of ASI under contract LONO 500172 and of  MURST via special programs "giovani ricercatori" Dip. Fisica Teorica Univ. Torino. 

{  
}
\vfill \eject

\newpage
{\bf Figures Caption}

1) Scheme for the generation of the double entangled photon pairs.  A Mach-Zender interferometer creates a state of the pump photon which is given by the superposition of the states corresponding to the photon following the long and the short path respectively. The pump photon  then generates or a horizontally polarised pair in the first type I (NLC1) crystal either (after having been rotated by a $\lambda / 2$ wave plate) a vertically polarised one in the second type I (NLC2) crystal. The  parametric down conversions of the two crystals are then superimposed using an optical condenser with a hole drilled in the centre for leaving pass the pump undisturbed. The optical path of idler, signal and pump are arranged by means of compensator elements (C) for not introducing any delay among these (see ref. \cite{nos} for details). The superposition of the probability of generating a pair in the first or in the second crystal originates the polarisation entanglement.
    
2) The scheme for the reception apparatus of Alice and Bob. A prism, properly rotated, allows a polarisation selection. On each arm exiting the prism a Mach-Zender interferometer is inserted with a phase shift on the long arm which is suitably arranged by the observer.  Photo-detectors are denoted by an ellipse.
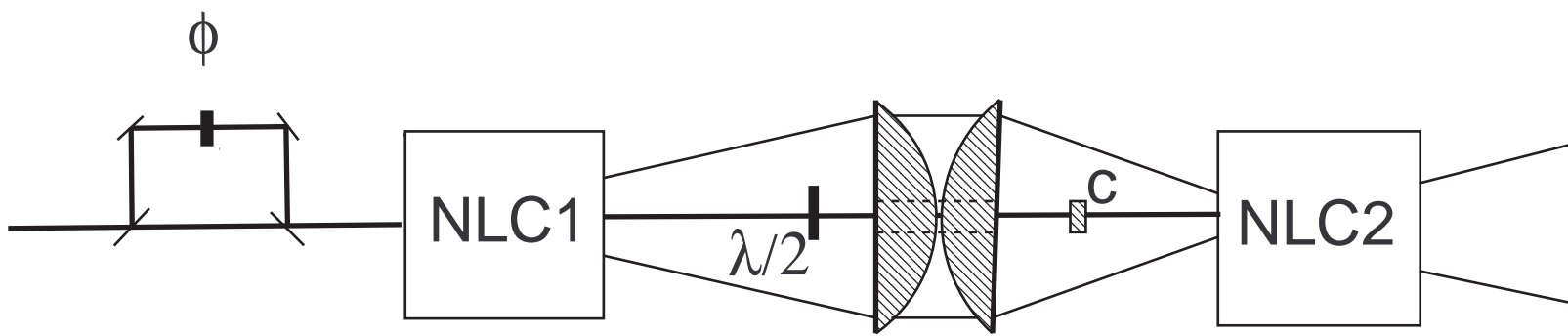
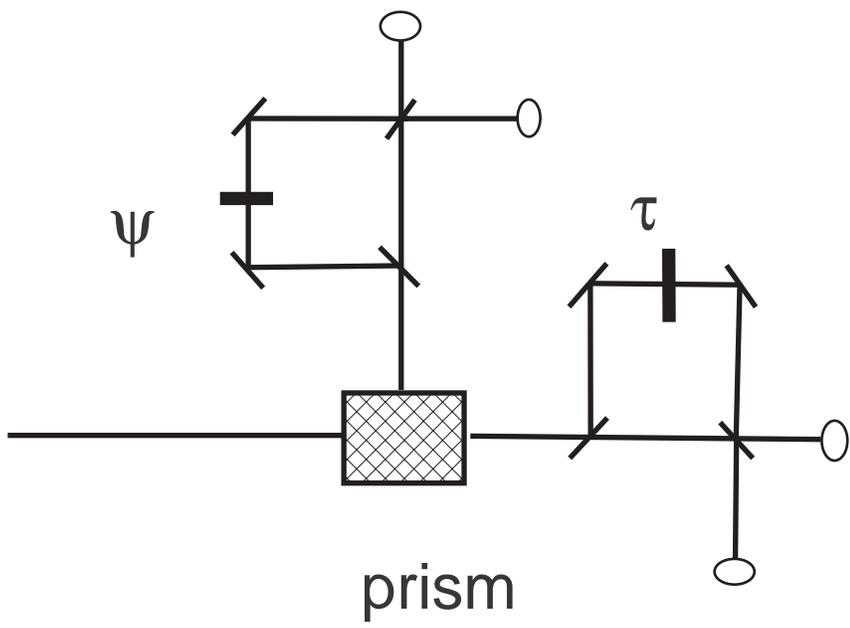
\end{document}